\pdfoutput=1

\documentclass[sigconf,screen=true,natbib,9pt,anonymous=false]{acmart}

\usepackage[moderate, leading=normal]{savetrees}

\copyrightyear{2018} 
\acmYear{2018} 
\setcopyright{acmlicensed}
\acmConference[ADCS '18]{23rd Australasian Document Computing Symposium}{December 11--12, 2018}{Dunedin, New Zealand}
\acmBooktitle{23rd Australasian Document Computing Symposium (ADCS '18), December 11--12, 2018, Dunedin, New Zealand}
\acmPrice{15.00}
\acmDOI{10.1145/3291992.3291993}
\acmISBN{978-1-4503-6549-9/18/12}

\usepackage{booktabs} 
\usepackage{graphicx}
\usepackage{url}
\usepackage{microtype}
\usepackage{tabularx,amsmath}
\usepackage{amssymb}
\usepackage{multirow}
\usepackage{color}
\usepackage[normalem]{ulem}
\usepackage[english]{babel}
\usepackage{colortbl}
\usepackage[noend]{algorithmic}
\usepackage{algorithm}
\usepackage{paralist}
\usepackage{enumerate}
\usepackage{enumitem}
\usepackage{acronym}
\usepackage[skip=0pt]{caption}
\usepackage{setspace}
\usepackage{balance}
\usepackage{subfig}
\usepackage[np, autolanguage]{numprint}

\setcitestyle{square,comma,numbers,sort&compress,sectionbib}

\newbox\itembox
\def\itemlistlabel#1{#1\hfill}
\def\itemlist#1{\setbox\itembox=\hbox{#1}%
                \list{}{\labelwidth\wd\itembox
                             \leftmargin\labelwidth
                             \advance\leftmargin by\itemindent
                             \advance\leftmargin by\labelsep
                             \let\makelabel\itemlistlabel}}

\newcommand{\myparagraph}[1]{\vspace{0.3\baselineskip}\noindent{\textbf{#1}}.~}

\acrodef{IR}{Information Retrieval}
\acrodef{WAND}{Weak or Weighted AND}

\title{The Potential of Learned Index Structures\\for Index Compression}

\author{Harrie Oosterhuis}
\orcid{}
\affiliation{%
\institution{University of Amsterdam}
}
\email{oosterhuis@uva.nl}

\author{J. Shane Culpepper}
\orcid{0000-0002-1902-9087}
\affiliation{%
\institution{RMIT University}
}
\email{shane.culpepper@rmit.edu.au}

\author{Maarten de Rijke}
\orcid{0000-0002-1086-0202}
\affiliation{%
\institution{University of Amsterdam}
}
\email{derijke@uva.nl}

\begin{document}

\begin{abstract}
Inverted indexes are vital in providing fast key-word-based search.
For every term in the document collection, a list of identifiers of
documents in which the term appears is stored, along with auxiliary
information such as term frequency, and position offsets.
While very effective, inverted indexes have large memory requirements
for web-sized collections.
Recently, the concept of learned index structures was introduced,
where machine learned models replace common index structures such as
B-tree-indexes, hash-indexes, and bloom-filters.
These learned index structures require less memory, and can be
computationally much faster than their traditional counterparts.
In this paper, we consider whether such models may be applied to
conjunctive Boolean querying.
First, we investigate how a learned model can replace document
postings of an inverted index, and then evaluate the compromises such
an approach might have.
Second, we evaluate the potential gains that can be achieved in terms
of memory requirements.
Our work shows that learned models have great potential in inverted
indexing, and this direction seems to be a promising area for future
research.
\end{abstract}

% CSS concepts are provided on submission page instead
%\begin{CCSXML}
%<ccs2012>
%<concept>
%<concept_id>10002951.10003317.10003338.10003343</concept_id>
%<concept_desc>Information systems~Learning to rank</concept_desc>
%<concept_significance>500</concept_significance>
%</concept>
%</ccs2012>
%\end{CCSXML}
%
%\ccsdesc[500]{Information systems~Learning to rank}
%%
%\keywords{Learning to rank; Online learning; Gradient descent}

\maketitle

\acresetall

% !TEX root = ./adcs2018-potential-learned-indexing.tex

\section{Introduction}
\label{sec:intro}

Search engines make large collections of documents accessible to
users, who generally search for documents by posing key-word based
queries.
For the best user experience, the user should be presented relevant
results as quickly as possible.
Inverted indexes allow systems to match documents with key-words in an
efficient manner~\cite{moffat1996self, zobel2006inverted}.
Due to their scalability, inverted indexes form the basis of most
search engines that cover large document collections.
They store inverted lists of the terms contained in each document in
the collection; for a given term, an inverted list stores a list with
all the documents in which it occurs.
%By traversing these lists one can easily retrieve the set of
%documents that contain \emph{any} of the words present in a query.
%For many relevance signals the union of the inverted lists contains
%all non-zero valued query-document pairs. 
%For instance, the scores of TF-IDF or BM25 are always zero if none of the query keywords appear in the document.
%Thus inverted indexes can be used to quickly determine the subset
%of documents that have to be considered for
%scoring~\cite{broder2003efficient}.
An important search operation performed on these lists is conjunctive
Boolean intersection, as it is routinely used in search engines for
early stage
retrieval~\cite{asadi2013effectiveness,clarke2016assessing,goodwin2017bitfunnel}
and for vertical search tasks such as product or job
search~\cite{russell2dsearch}.
Boolean queries are computed by intersecting the inverted lists of
the query terms~\cite{culpepper2010efficient}, and the result set
typically includes documents that contain \emph{all} of the query
terms, {\em including} the stopwords.

Despite consistent advances in compressed index representations over
the years~\cite{t14-adcs,ov14sigir,lemire2015decoding}, the cost of
storing all relevant data (such as stopword data) to effectively
search sizable collections can be a bottleneck in scaling to
increasingly larger collections.
%For large collections, many terms appear in hundreds of thousands of
%documents, the stored list for these terms will be proportionally
%large.
%Several approaches have been introduced to deal with the memory
%burdens of inverted indexes.
%Compression techniques can reduce storage space, but adds
%computational costs during retrieval when decompression has to be
%performed~\cite{lemire2015decoding}.
One interesting alternative is to use a bitvector to store the
document vector of high frequency
terms~\cite{moffat2007hybrid,kane2014skewed}.
%Since users only consider a small number of the top-retrieved documents, a common approach is to discard documents with a low term-frequency from inverted lists.
However, there are limits to what compressing exact representations
can achieve.
%These so called champion-lists only contain the documents with the
%highest expected relevance for a term, i.e.
%the highest term-frequency, or highest impact score for a relevance
%signal~\cite{broder2003efficient, ding2011faster}.
%This greatly reduces the size of an inverted index, however, the
%results for queries are no longer guaranteed to be correct.
%For instance, a document may incorrectly not appear in a conjunctive
%queries because it happened to fall outside of the champion-list of
%one the query terms.
%Unfortunately there is no existing method that decreases storage like
%champion-lists while maintaining guarantees about query-results.

Recently the idea of learned index structures has been proposed by
\citeauthor{kraska2018case}~\cite{kraska2018case}.
Here, machine learned models are optimized to replace common index
structures such as B-tree-indexes, hash-indexes, and bloom-filters.
The benefit of learned index structures is that they require
less memory and can be evaluated substantially faster than their
traditional counterparts.

In this study we examine whether learned index structures can be
used to reduce space in a Boolean search scenario, and investigate
the effect this would have on exactness guarantees an index can
provide.
Using existing document collections, we estimate the space savings
that such an approach could achieve.
Our results show that a learned index structure approach has the
potential to significantly reduce storage requirements, and still
provide performance guarantees.
%are similar to that of just using champion-lists.
%Conversely, the learned model provides more functionality as it can
%estimate matches between terms and documents outside of a
%champion-list.
%Thus it seems the learned model approach is very fruitful direction
%for future research.
%
The research questions we address are:

\smallskip

\begin{enumerate}[align=left, label={\bf RQ\arabic*}, leftmargin=*, nosep, topsep=5pt, itemsep=3pt,nosep]
\item How might learned indexes be used to support search based on Boolean intersection?
\label{rq:possible}
\item Would learned indexes provide any space benefits over current compressed
indexing schemes?\label{rq:gain}
\end{enumerate}

% !TEX root = adcs2018-potential-learned-indexing.tex

\section{Related Work}
\label{sec:relatedwork}
%
%\shane{Sketch of idea
%\begin{itemize}
%\item Conjunctive Boolean queries require only document identifiers, but
%also must include all terms, like stopwords.
%\item The operation is simple, take the shortest list and iteratively
%search for each item in the other term lists using a one-sided binary
%finger search (galloping search).
%\item Technical problem for us -- what if the document posting is ``learned''?
%So if two common lists must be searched, we cannot do it. For this we
%include a top-$k$ list for the learned lists.
%\item There might be other tricks like a ``dirty bit'' vector which identifies
%regions containing terms.
%\item If we assume that we are in a multistage scenario, we only need to 
%guarantee some maximum number of items.
%\end{itemize}
%}

Boolean intersection has been a fundamental component of information
retrieval systems for more than fifty years.
In fact, early search systems were entirely reliant on Boolean
retrieval models~{\cite{croft2010search}}.
In recent years, ranked retrieval models have become more important,
but Boolean operations are still a fundamental component in a variety
of search tasks~\cite{goodwin2017bitfunnel,russell2dsearch}.

One important application of Boolean intersection is as either a
feature in multi-stage retrieval~\cite{msoh13acmtois}, or as a
filtering stage in a multi-stage
pipeline~\cite{p10-query,goodwin2017bitfunnel}.
In all of these, the key idea is to apply expensive feature
extraction and machine learning models on a subset of the most promising
candidate documents to ensure early-precision is maximized in the
final result set~\cite{wlm11sigir,cgbc17-sigir}.

Machine learning has been applied in early-stage retrieval to predict
the number of documents to pass through to the next
stage~{\cite{ccl16-adcs}} and even to predict which top-$k$
processing algorithm should be used per query~{\cite{mc+18-wsdm}}.
But in current cascaded and multi-stage retrieval models the use of
machine learning algorithms is often deferred to later stages of the
retrieval process as traditionally such algorithms were not optimized
for efficiency.
The recent introduction of learned index structures by
{\citet{kraska2018case}} is changing that perception.
They have shown that common index structures such as B-tree-indexes,
hash-indexes, and Bloom-filters can be replaced by learned models,
while bringing both gains in memory and computational costs.
Moreover, by applying recursive models a learned index structure can
fallback on traditional structures for sub-cases where a learned
model performs poorly.
Consequently, learned models can provide the same correctness
guarantees as their traditional counterparts.
Given the potential advantages of learned models, we explore this
line of research in a constrained early-stage retrieval scenario --
specifically, can a learned indexing representation be used for
conjunctive Boolean retrieval?
If so, what are the performance implications?

% !TEX root = adcs2018-potential-learned-indexing.tex

\begin{algorithm}[tb]
\caption{The Exhaustive Iterative Approach.} 
\label{alg:iterative}
\begin{algorithmic}[1]
\STATE $q_1,\ldots,q_n \leftarrow \textit{receive\_query}$ 
\STATE $r \leftarrow []$
\FOR{$d \in D$}
     \IF{$\forall q_i \in [q_1, \ldots, q_n], \quad f(q_i, d) = 1$}
        \STATE $\textit{append}(r, d)$
    \ENDIF
\ENDFOR
\RETURN r
%	\hfill \textit{\small // update the ranking model} \label{line:novel:update}
\end{algorithmic}
\end{algorithm}

\begin{algorithm}[tb]
\caption{The Two-Tiered Approach.} 
\label{alg:topdoc}
\begin{algorithmic}[1]
\STATE $q_1,\ldots,q_n \leftarrow \textit{receive\_query}$ 
\STATE $l_1,\ldots,l_n \leftarrow \textit{truncated\_lists\_for\_terms}(q_1,\ldots,q_n)$
\STATE $L \leftarrow \bigcup_{i = 1}^{n} l_i$
\STATE $r \leftarrow []$
\FOR{$d \in L$}
     \IF{$\forall q_i \in [q_1, \ldots, q_n], \quad f(q_i, d) = 1$}
        \STATE $\textit{append}(r, d)$
    \ENDIF
\ENDFOR
\RETURN r
\end{algorithmic}
\end{algorithm}

\begin{algorithm}[tb]
\caption{The Block Based Approach.} 
\label{alg:block}
\begin{algorithmic}[1]
\STATE $q_1,\ldots,q_n \leftarrow \textit{receive\_query}$ 
\STATE $b_1,\ldots,b_n \leftarrow \textit{block\_lists\_for\_terms}(q_1,\ldots,q_n)$
\STATE $B \leftarrow \bigcap_{i = 1}^{n} b_i$
\STATE $r \leftarrow []$
\FOR{$b \in B$}
    \FOR{$d \in \textit{document\_range\_of\_block}(b)$}
        \IF{$\forall q_i \in [q_1, \ldots, q_n], \quad f(q_i, d) = 1$}
            \STATE $\textit{append}(r, d)$
        \ENDIF
    \ENDFOR
\ENDFOR
\RETURN r
\end{algorithmic}
\end{algorithm}

\begin{figure*}[t]
\centering
\caption{Top: the distribution of document frequencies. Bottom: the minimum number of terms that appear at different fractions of the compressed inverted index. From left to right: results for the Robust, GOV2 and ClueWeb collections. }
\includegraphics[width=0.32\textwidth]{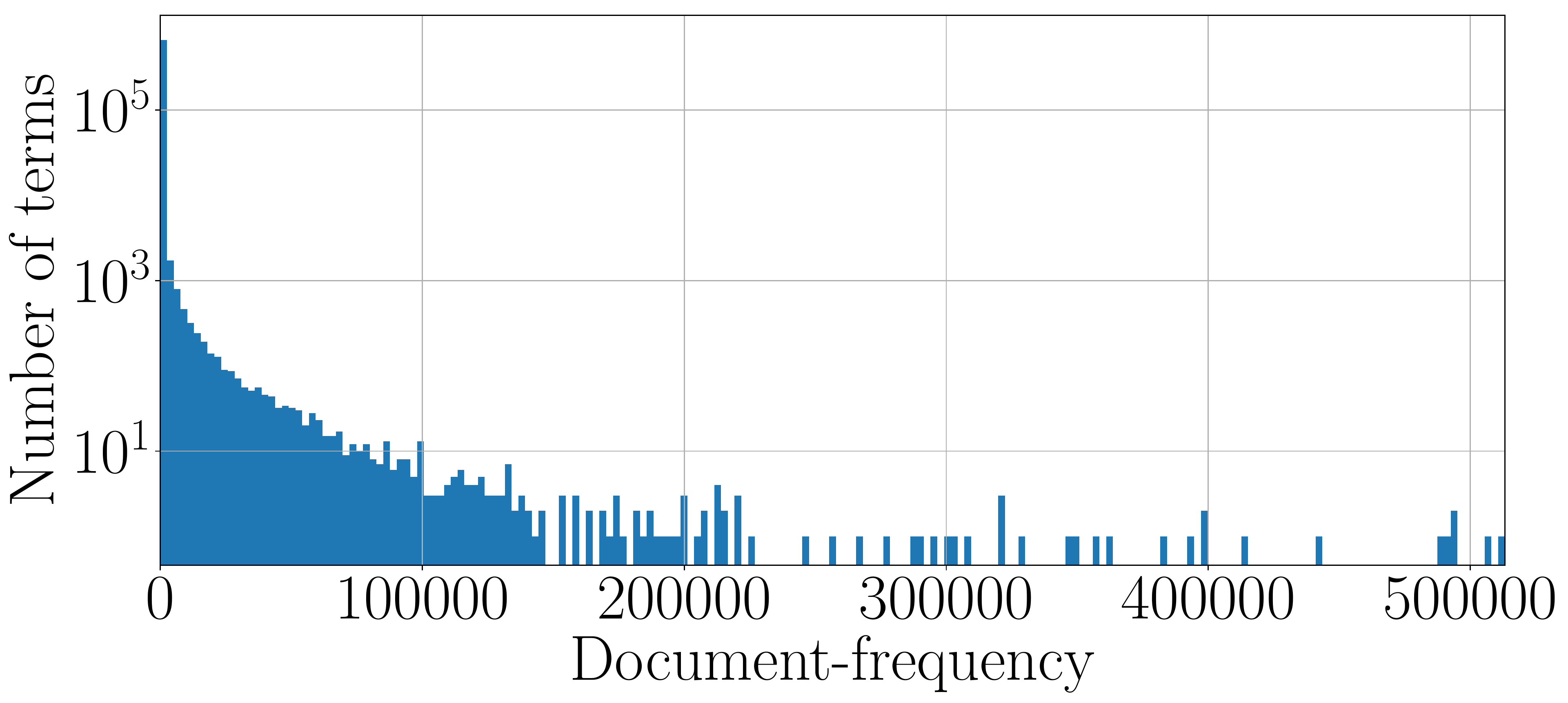}
\includegraphics[width=0.32\textwidth]{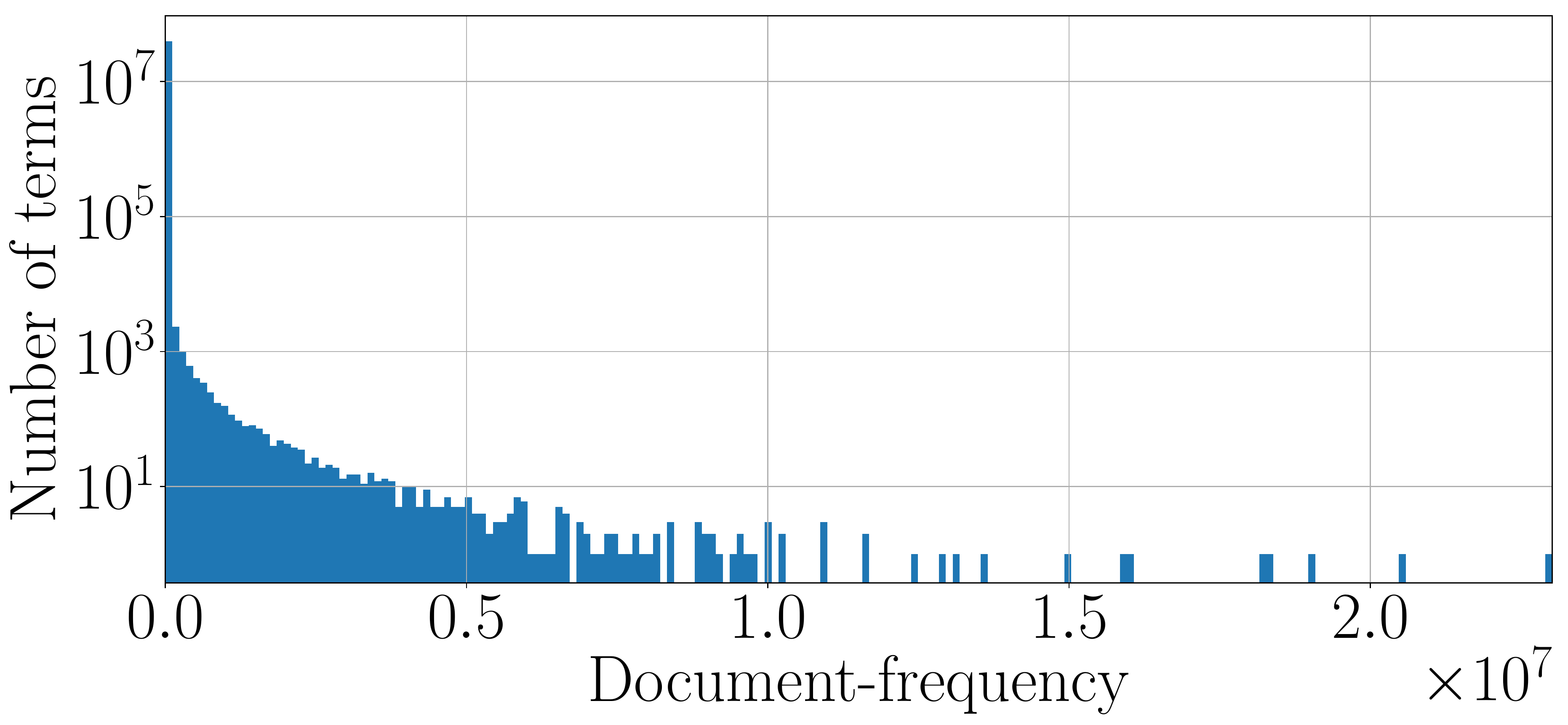}
\includegraphics[width=0.32\textwidth]{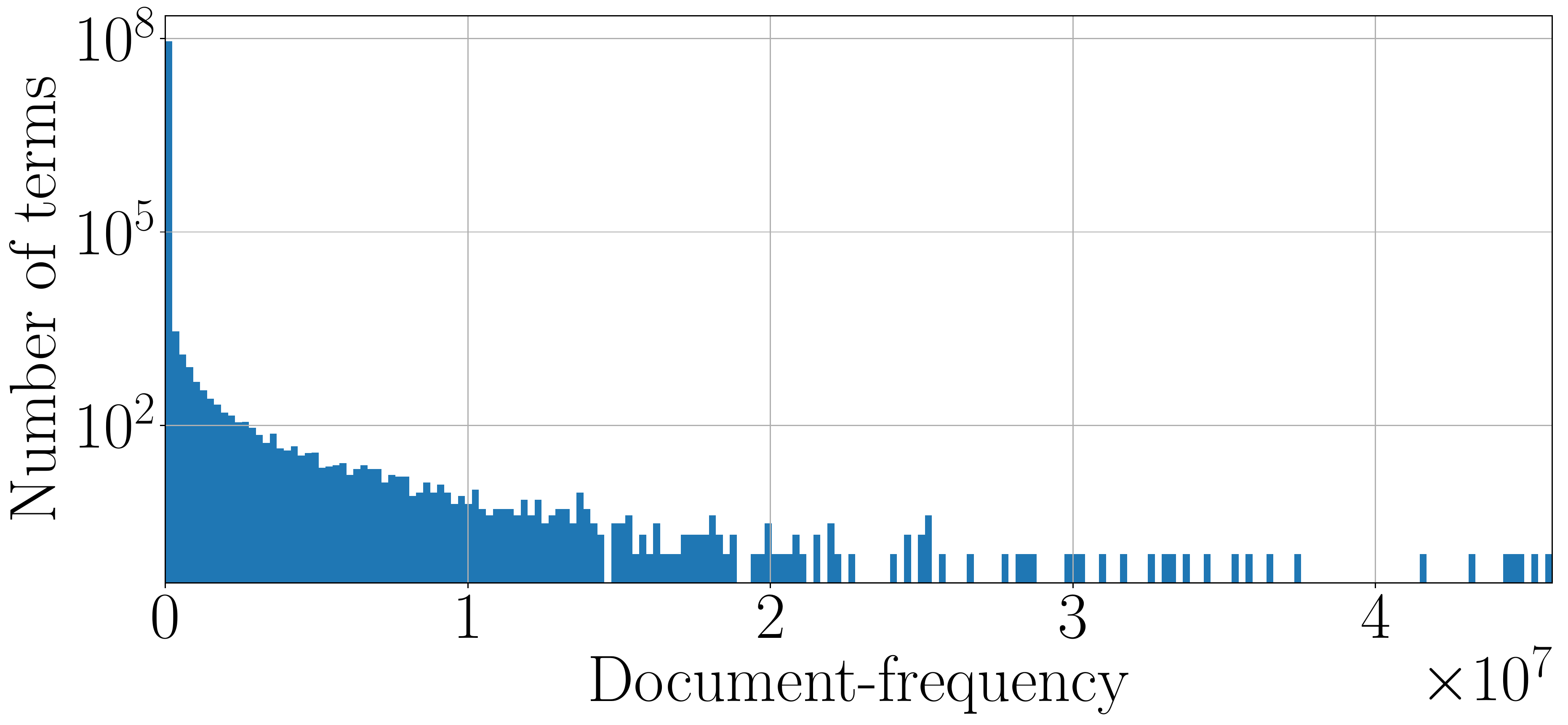}\\
\includegraphics[width=0.32\textwidth]{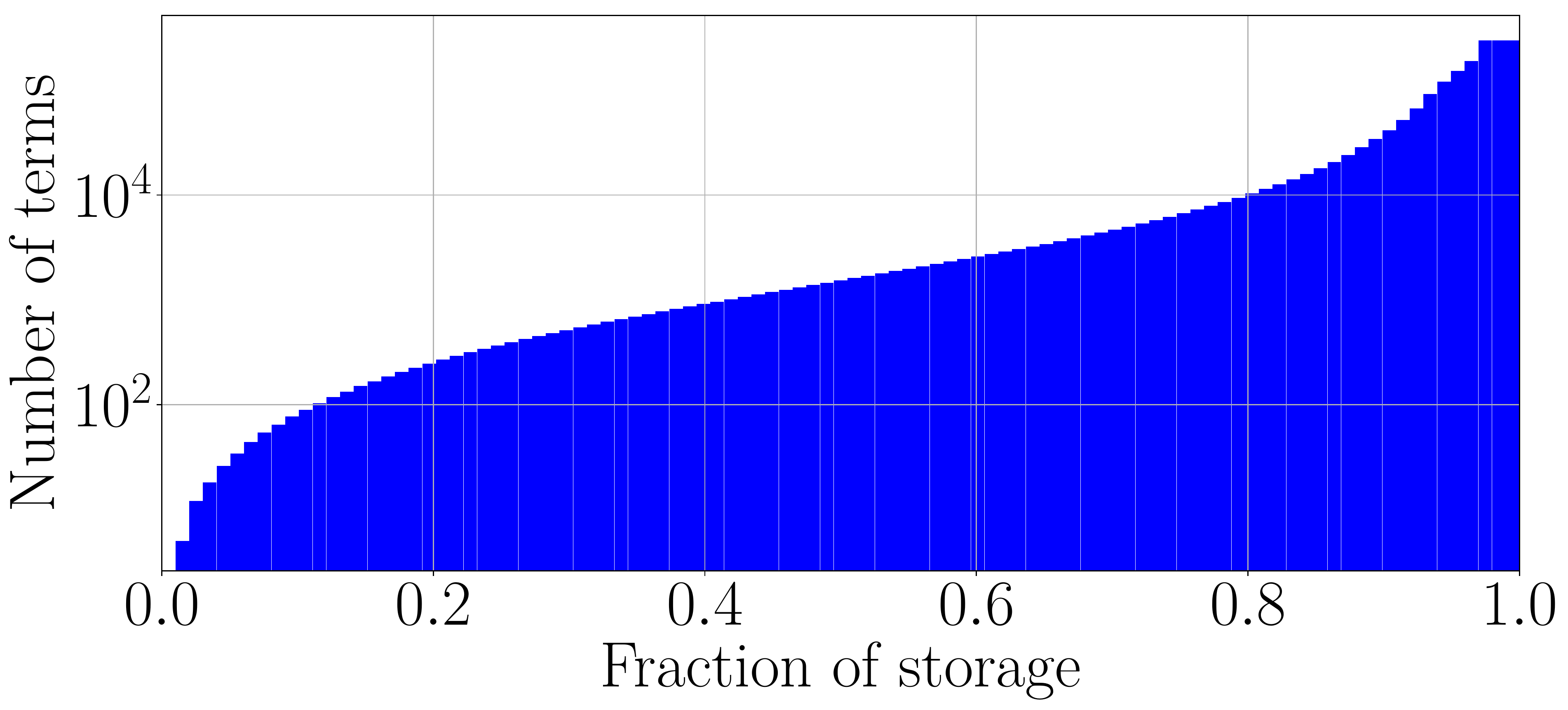}
\includegraphics[width=0.32\textwidth]{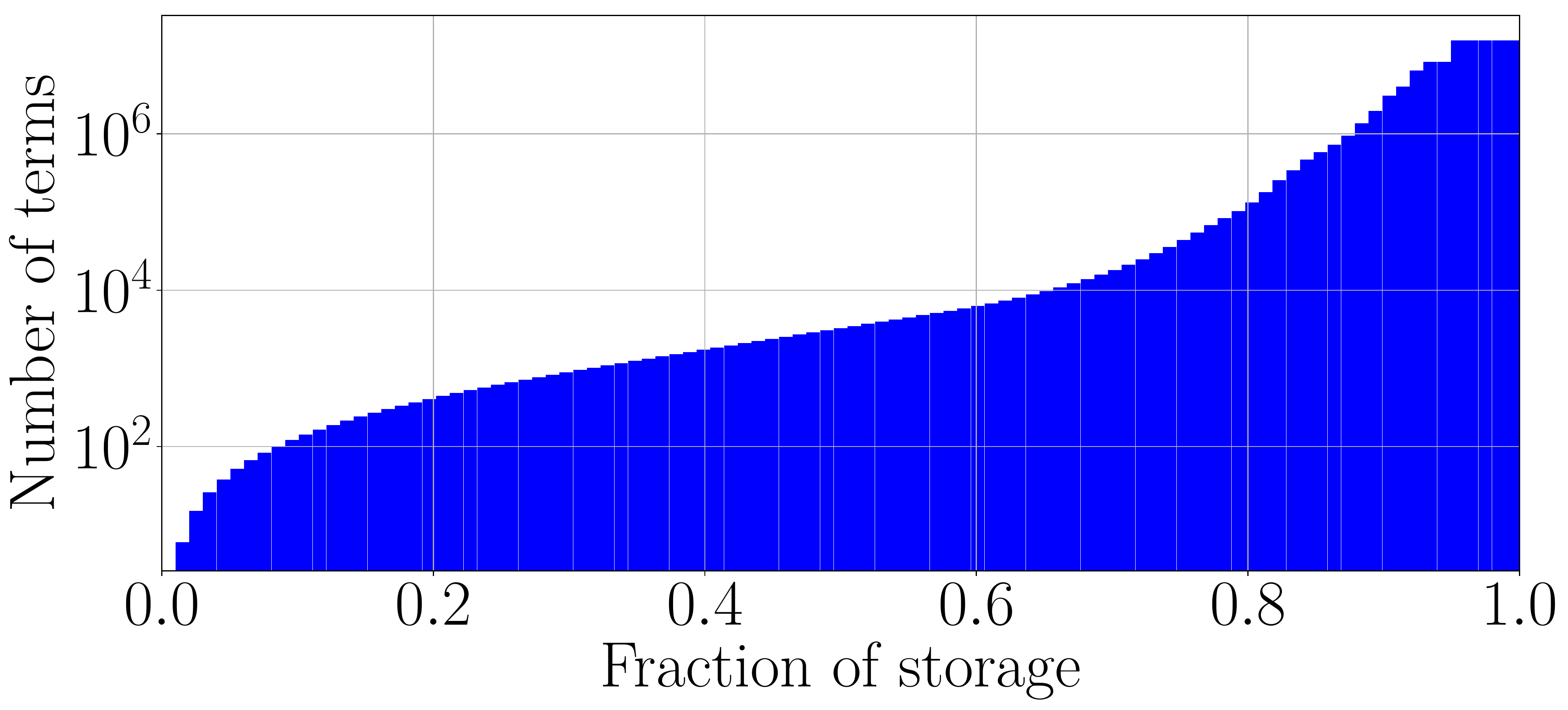}
\includegraphics[width=0.32\textwidth]{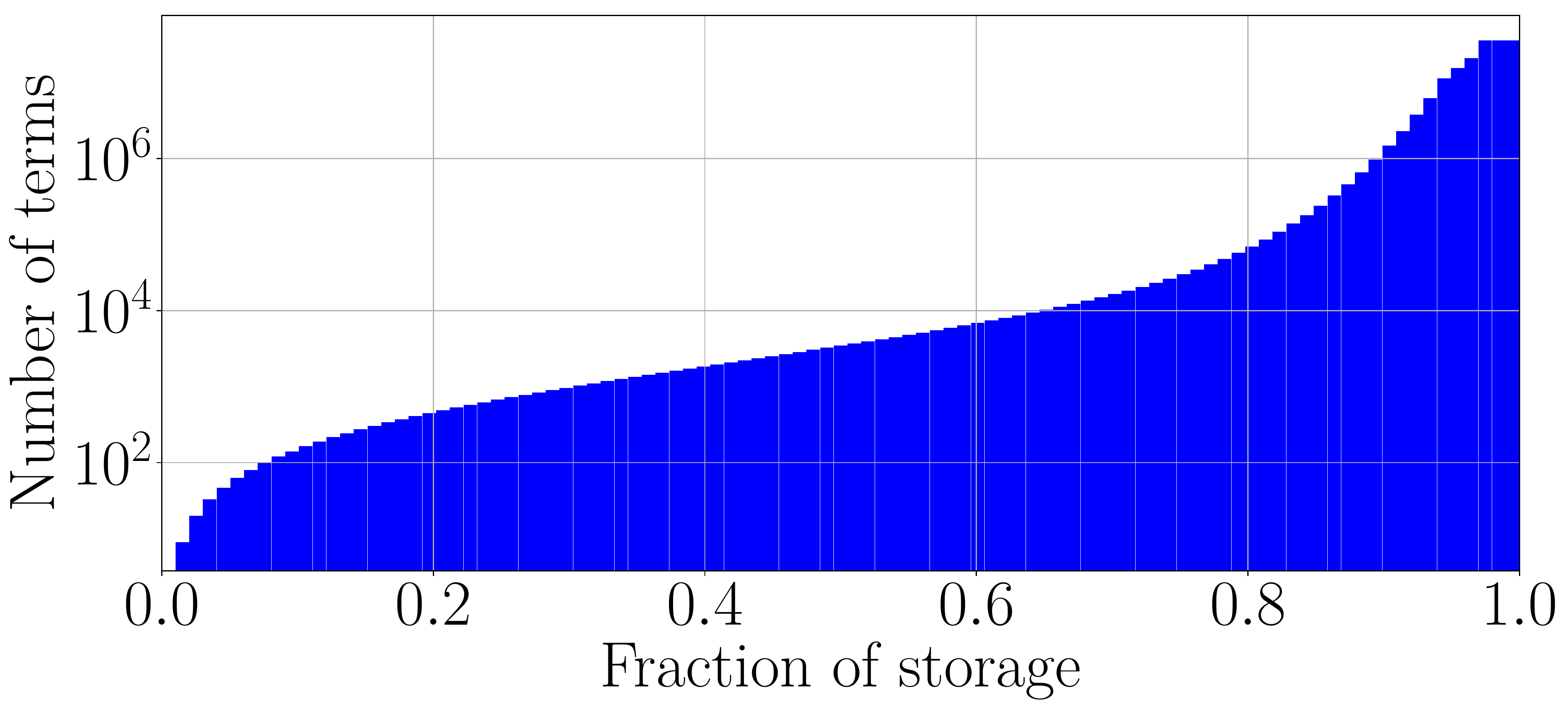}
\label{fig:freqcompr}
\vspace{-1\baselineskip}
\end{figure*}

\section{Applicability for Inverted Indexes}
\label{sec:applicability}

In this section we answer \ref{rq:possible}: in what ways a learned
index model can support Boolean intersection based search.

%Ideally a learned model should produce an inverted lists instantly and without large storage requirements.
%Such a function would take a term $t$ and produce the inverted list, i.e. $f(t) = [d_1,d_2,\dots]$.
%While there are existing deep-learning models that produce sequences~\cite{sutskever2014sequence}, none have been applied to the scale of document collections or for similar purposes.
%As a result, we will assume that such a generative approach will not work for inverted indexes.

We will consider models that act as learned Bloom-filters because
they have commonly been applied to conjuctive Boolean problems.
Moreover, \citeauthor{kraska2018case}~\cite{kraska2018case} have shown
that they can be applied to sizeable datasets and outperform
traditional Bloom-filters in both speed and memory requirements.
However, learned index structures have only been optimized for tasks
involving a single set~\cite{kraska2018case}.
In contrast, each document in an inverted index could be seen as an
individual set, thus making the problem substantially more complex.

For this study we will assume that a function $f(t,d) \in \{0,1\}$ can be
learned perfectly so that for a term $t$ and document $d$:
\begin{align}
f(t,d) = 
    \begin{cases}
    1 & t \in d, \\
    0  & t \not \in d.
    \end{cases}
\end{align}
In theory any deep neural network that is expressive enough could be
optimized for an entire document collection without any errors.
In practice, such a model has to be relatively sizeable and requires
a very long period of optimization.
Therefore, one may choose to compromise some correctness for
practical reasons.
In this study, we will not discuss the specifics of such a model or
its optimization and instead focus on how it could be applied.

\subsection{Exhaustive Iterative Approach}

A straightforward approach to conjunctive Boolean functions using the
model $f$ would be to iterate over the entire document collection.
Algorithm~\ref{alg:iterative} displays what this approach could look
like.
It is clear that, per query, there is a huge computational cost
proportional to the number of documents in the collection.
However, this approach guarantees the correct results for conjunctive
Boolean queries.
Moreover, the only storage it requires is for the model $f$, thus it
can provide the biggest gains in memory by completely replacing an
inverted index.
In practice this approach will most likely be avoided because of its
computational costs, yet it provides an interesting example of how
the storage requirements could be completely minimized.

\subsection{Two-Tiered Approach}
\label{sec:topdoc}
The previous approach iterated over all documents in the collection, which has high computational costs.
An existing method of speeding up retrieval is to use two-tier retrieval~\cite{rossi2013fast}.
Here an index is divided in two partitions, one of which is of smaller size on which queries can be pre-processed quickly.
We also propose a two-tiered approach where an inverted index is split into a smaller partition with truncated lists, and a larger partition with the remainder of the lists.
We will assume that the size of the second partition is not important, but that the goal is to minimize the size of the first partition.
The first partition consists of the inverted lists of each term but truncated to length $k$, the remainder of each list appears in the second partition.
We will not make any assumptions about which parts of the lists are included in the truncations.
Then Algorithm~\ref{alg:topdoc} displays how one may use the learned model $f$ to search through the first partition.
This approach only has to iterate over the intersection of the truncated lists, thus it is computationally more efficient than the previous approach.
However, to retrieve all results the truncated lists will not always suffice.
If \emph{all} terms in a query have a document frequency greater than $k$ then results may be missing after passing over the first partition.
At this point, the algorithm could fallback by also considering the second partition.
Conversely, correctness is guaranteed if \emph{at least one} query-term appears in $k$ or less documents.
By applying the learned model $f$ there is no need to use the second partition here.
Thus for queries with at least one infrequent term the first partition and learned model are guaranteed to provide correct results.
%The two-tiered approach allows us to split an inverted index into one large and another substantially smaller partition, while being able to process a large part of queries only on the smaller partition.
This approach may be particularly advantageous when the smaller size allows the first partition to fit in memory components with faster access.

\subsection{Block Based Approach}

Lastly, we introduce an approach inspired by existing signature files
and partitioned approaches \cite{goodwin2017bitfunnel} in
Algorithm~\ref{alg:block}.
A document collection may be partitioned into multiple \emph{blocks},
each containing a subset of documents.
For every term, a list indicating the blocks in which their matching
documents appear is stored.
Then the intersection of the lists for every query-term provides
restricted ranges in which results for a conjunctive Boolean query
appear.
Finally, these ranges can be traversed with the learned model $f$ to
retrieve all documents for the conjunctive Boolean query.
The computational costs of this approach are limited by the size of
the partitions.
Reductions in storage can be achieved since only a list of partitions
has to be stored.
We note that for very infrequent terms, traditional inverted lists may
still be stored resulting in hybrid
presentations~\cite{moffat2007hybrid}.
In addition to the storage gains, this approach still guarantees
correct results for conjunctive Boolean queries.

\begin{figure*}[tb]
\centering
\caption{Top: The estimated upper and lower bounds in terms of storage space required. Bottom: The number of terms that need to be replaced. From left to right: results for the Robust, GOV2 and ClueWeb09B collections. }
\includegraphics[width=0.32\textwidth]{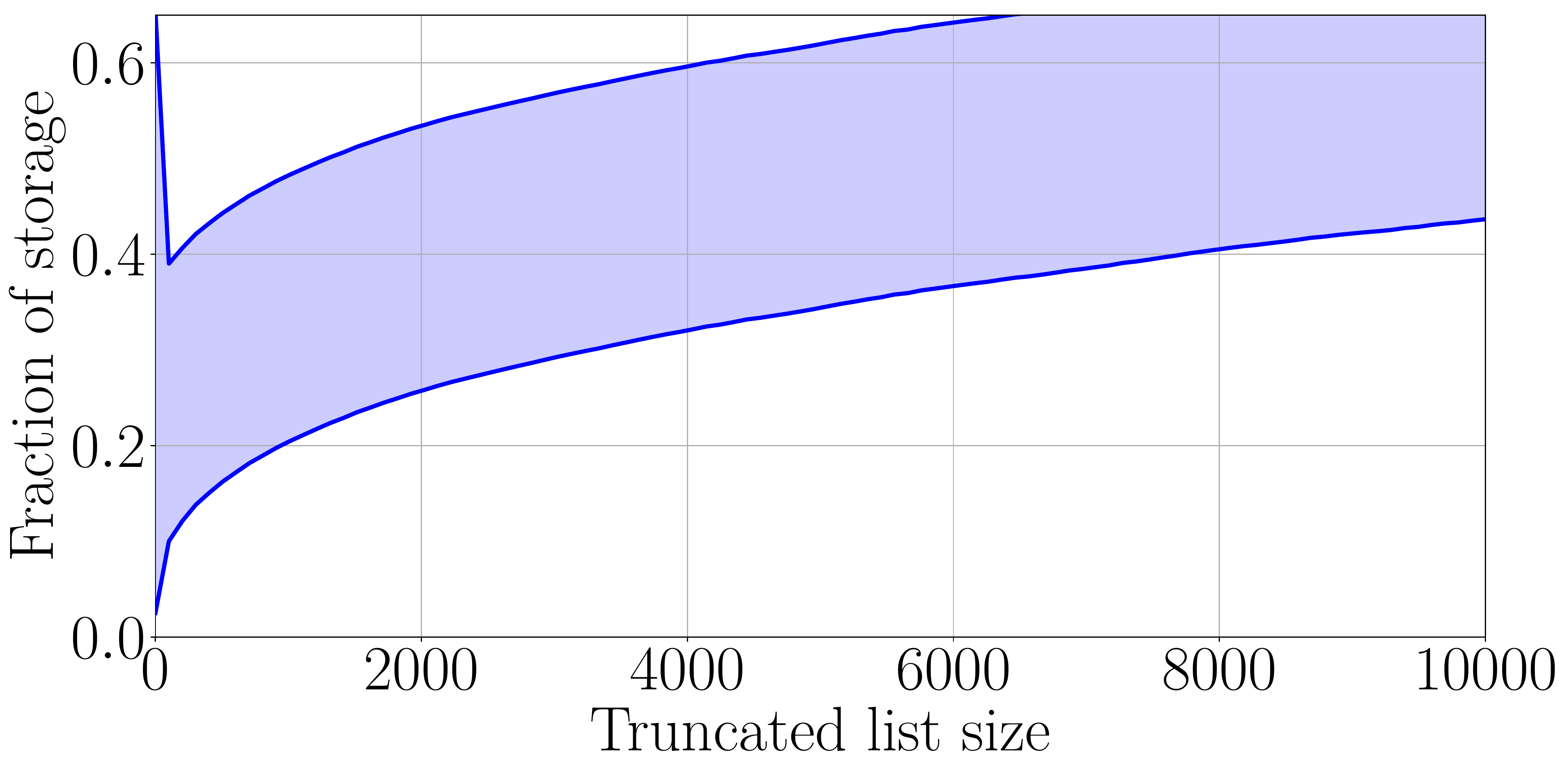}
\includegraphics[width=0.32\textwidth]{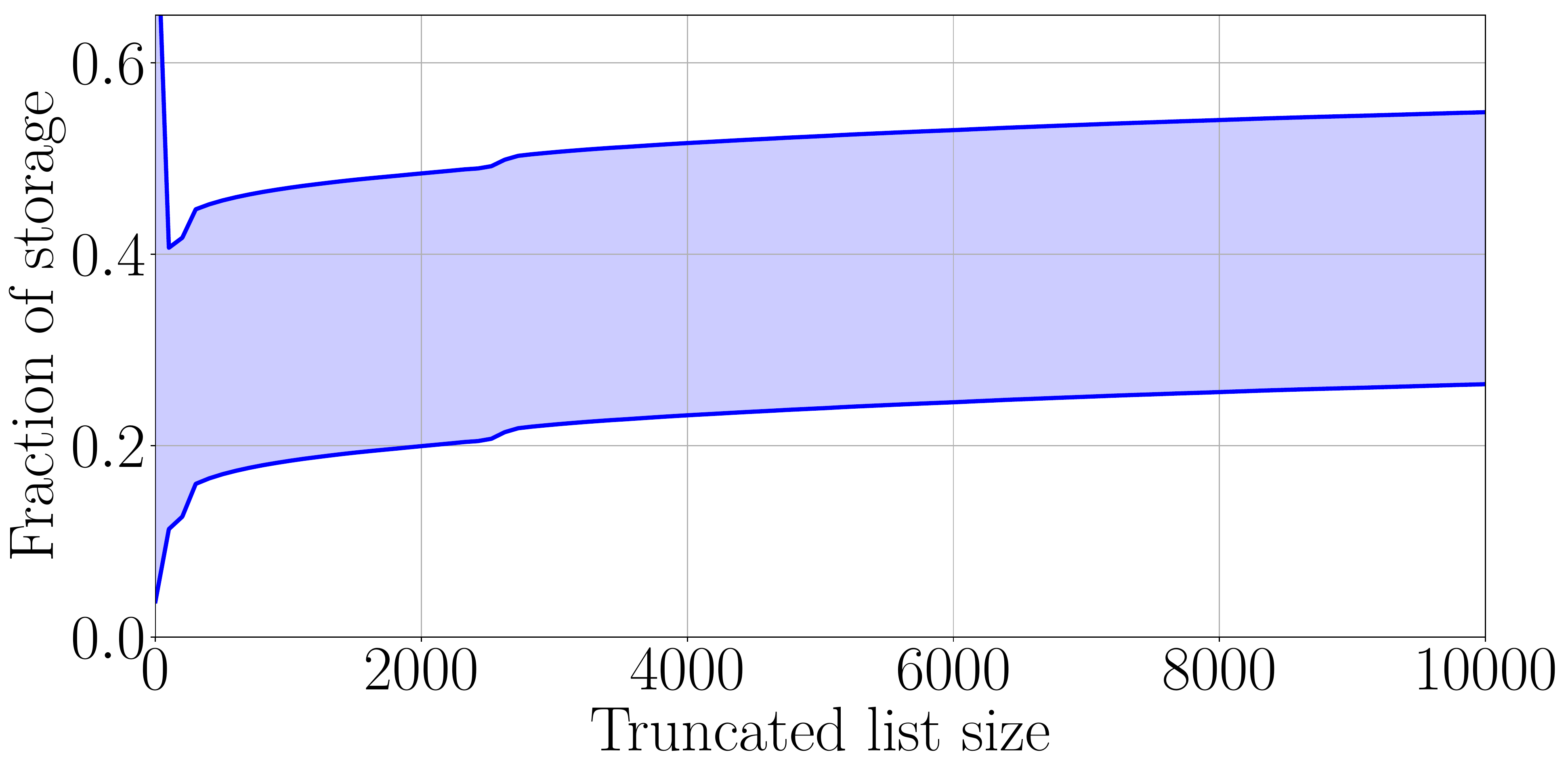}
\includegraphics[width=0.32\textwidth]{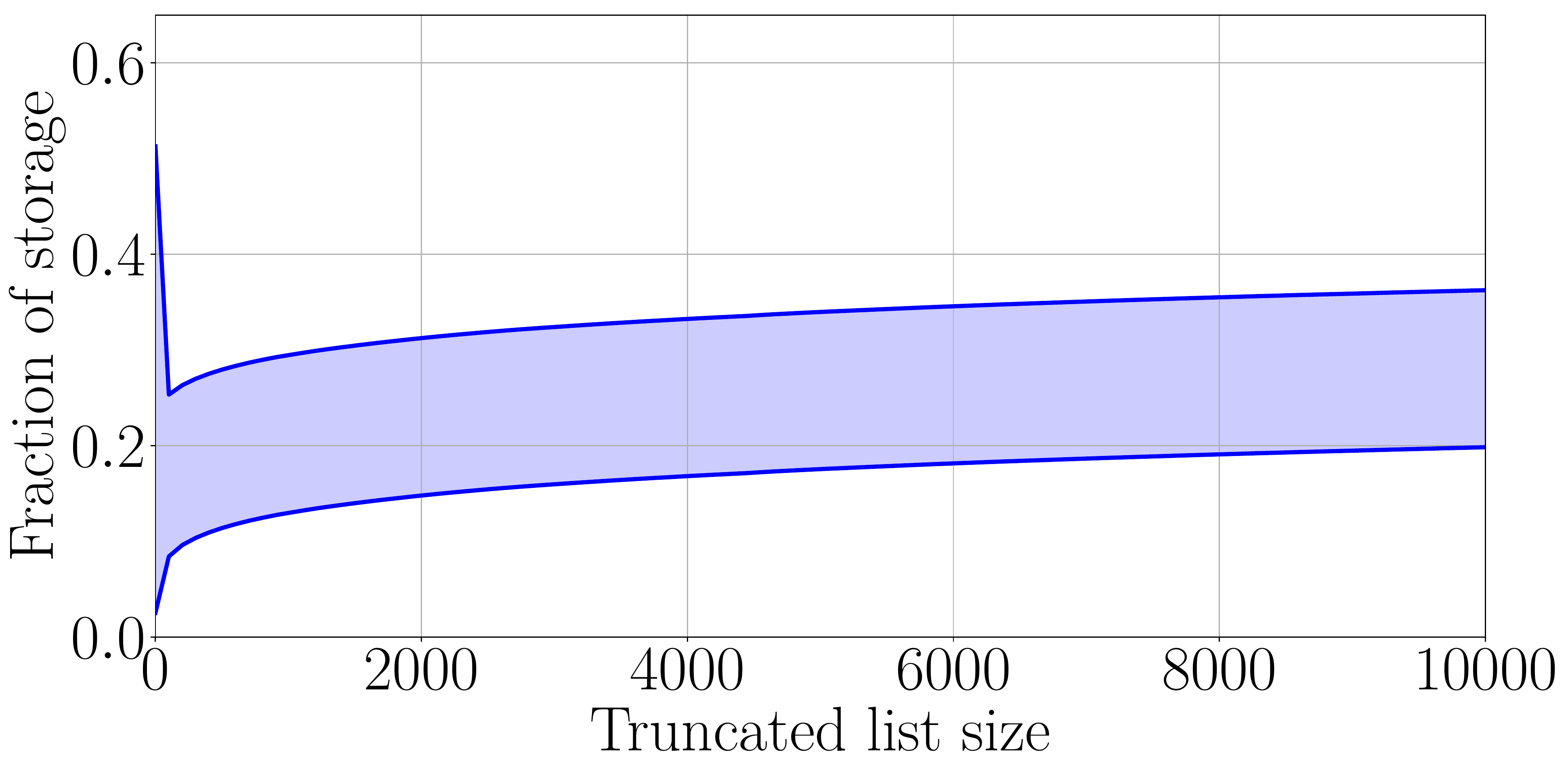}\\
\includegraphics[width=0.32\textwidth]{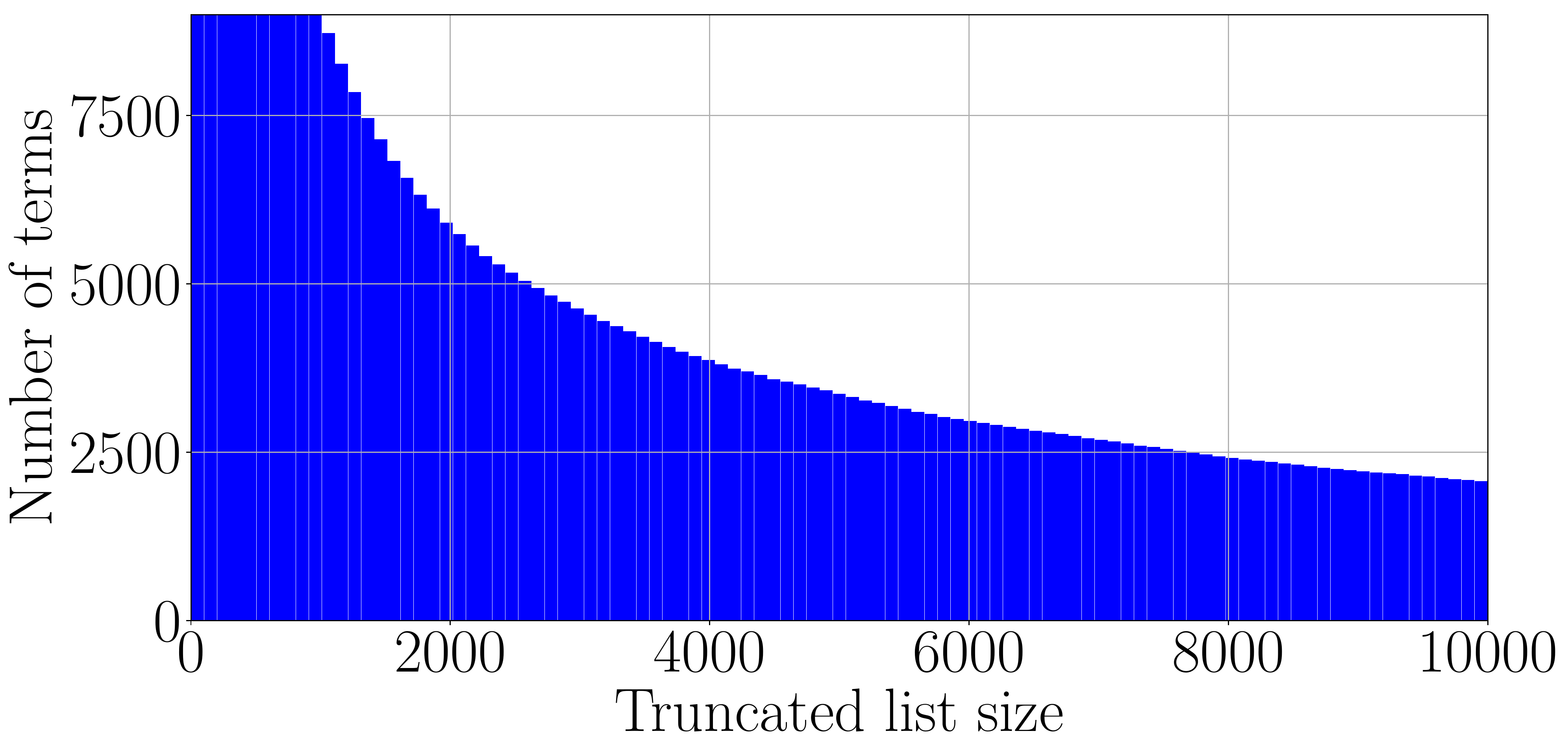}
\includegraphics[width=0.32\textwidth]{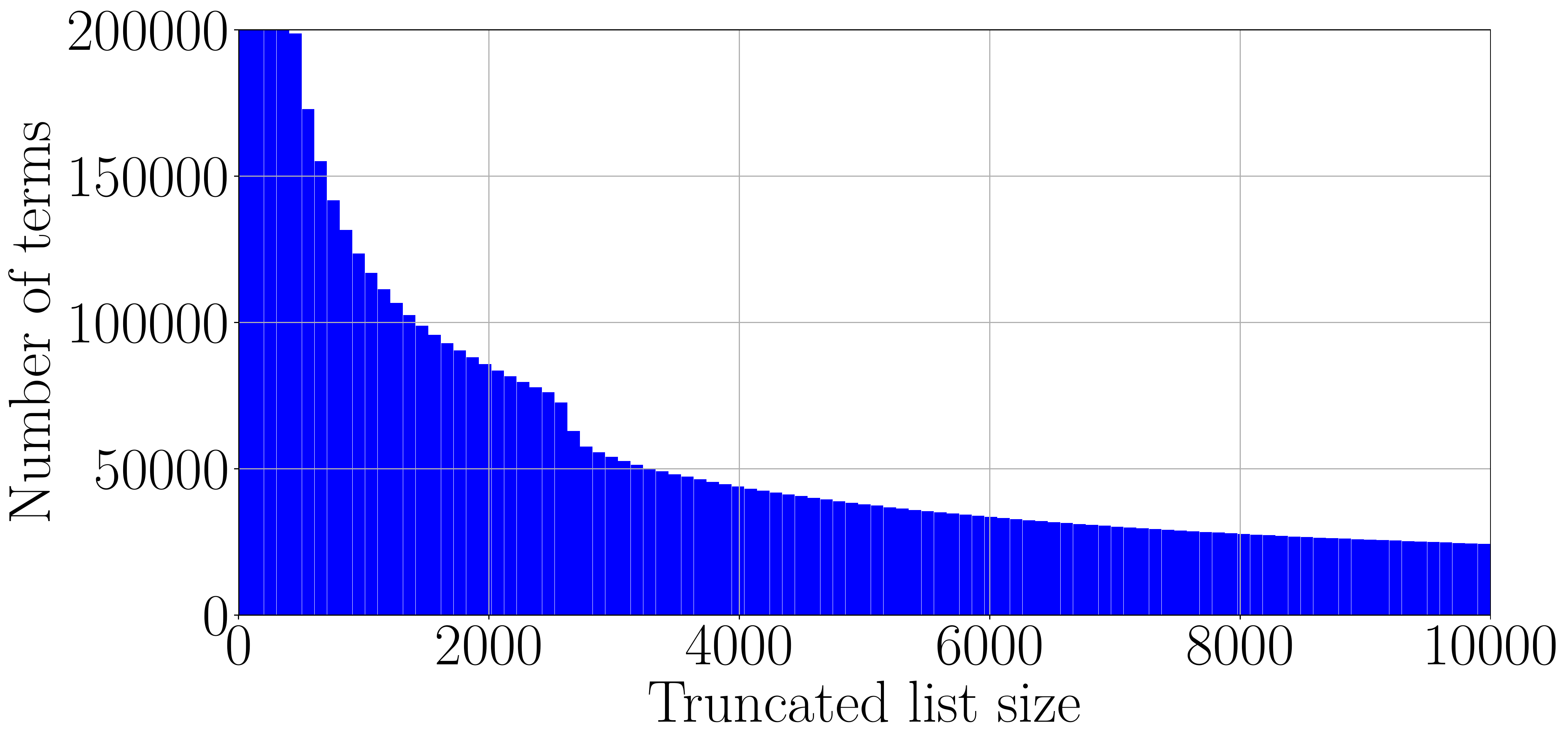}
\includegraphics[width=0.32\textwidth]{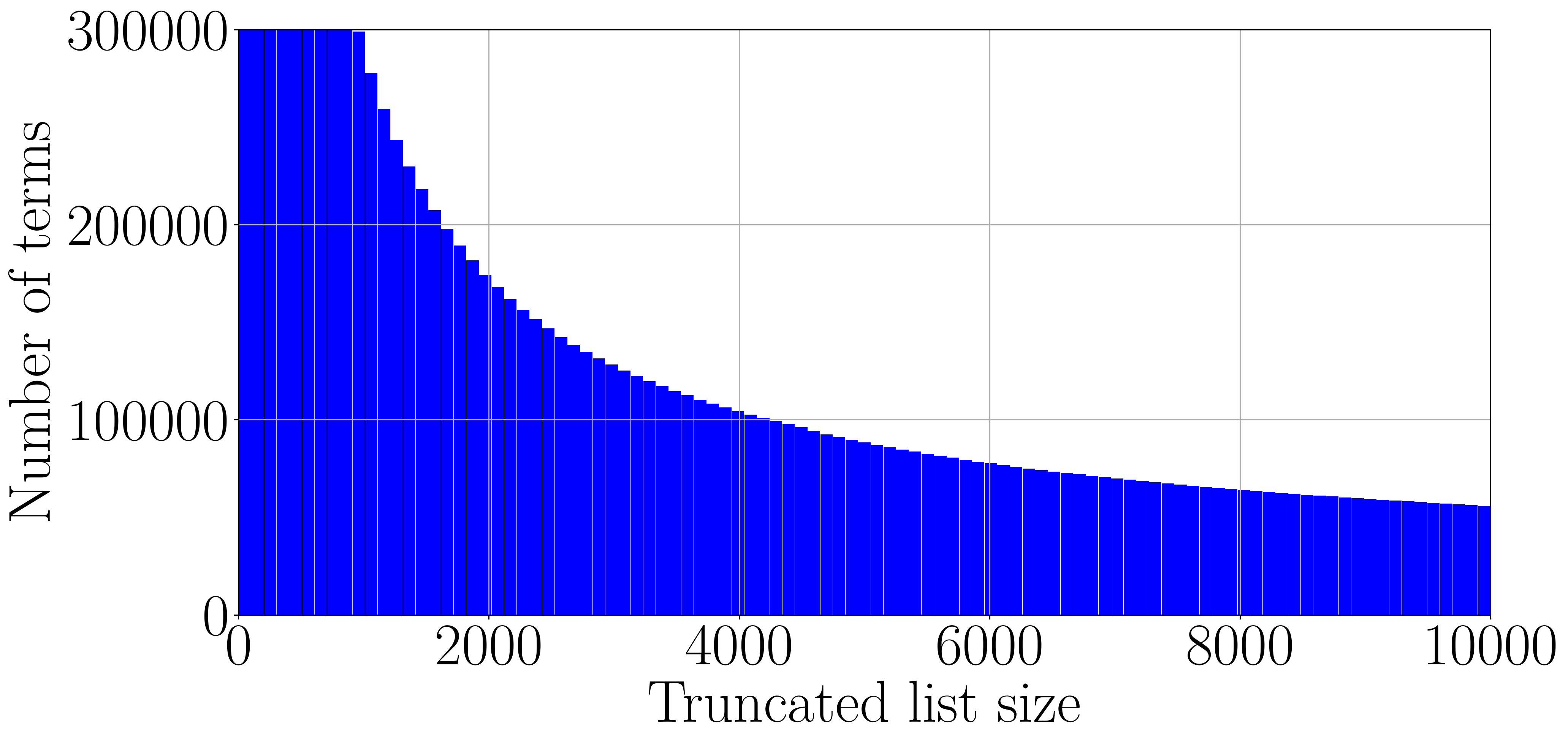}
\label{fig:bounds}
\vspace{-1\baselineskip}
\end{figure*}

\begin{figure*}[tb]
\centering
\caption{Percentage of queries with guaranteed correct results in the first-tier by varying truncated list sizes. From left to right: Robust, GOV2 and ClueWeb09B. }
\includegraphics[width=0.32\textwidth]{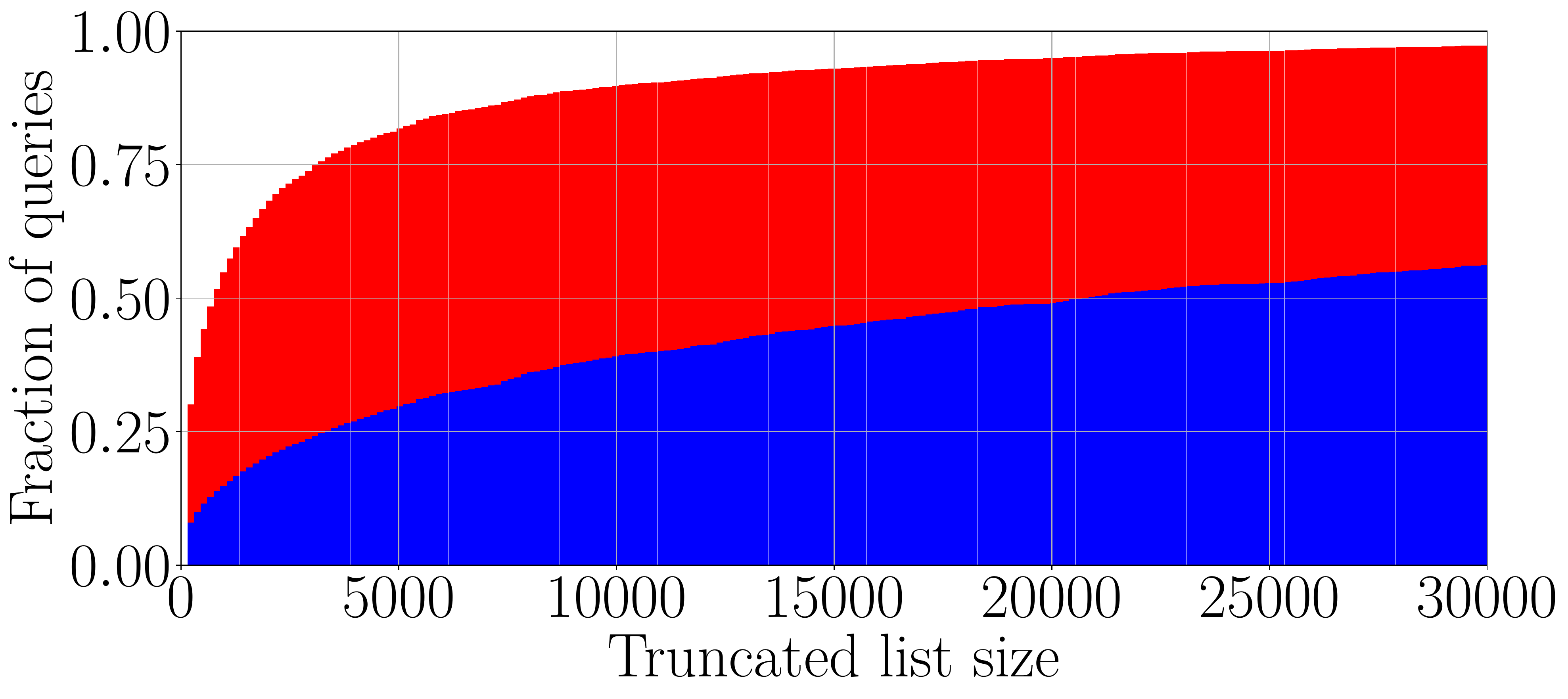}
\includegraphics[width=0.32\textwidth]{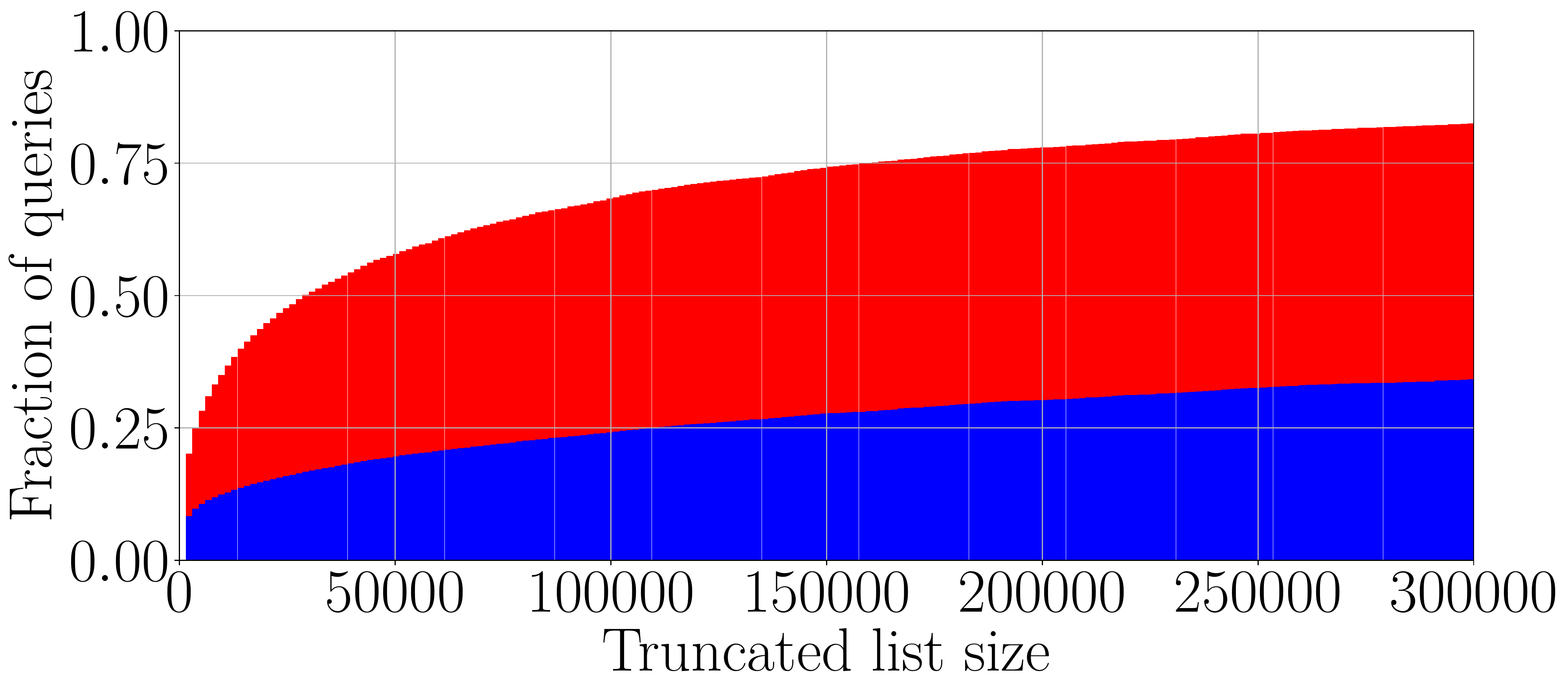}
\includegraphics[width=0.32\textwidth]{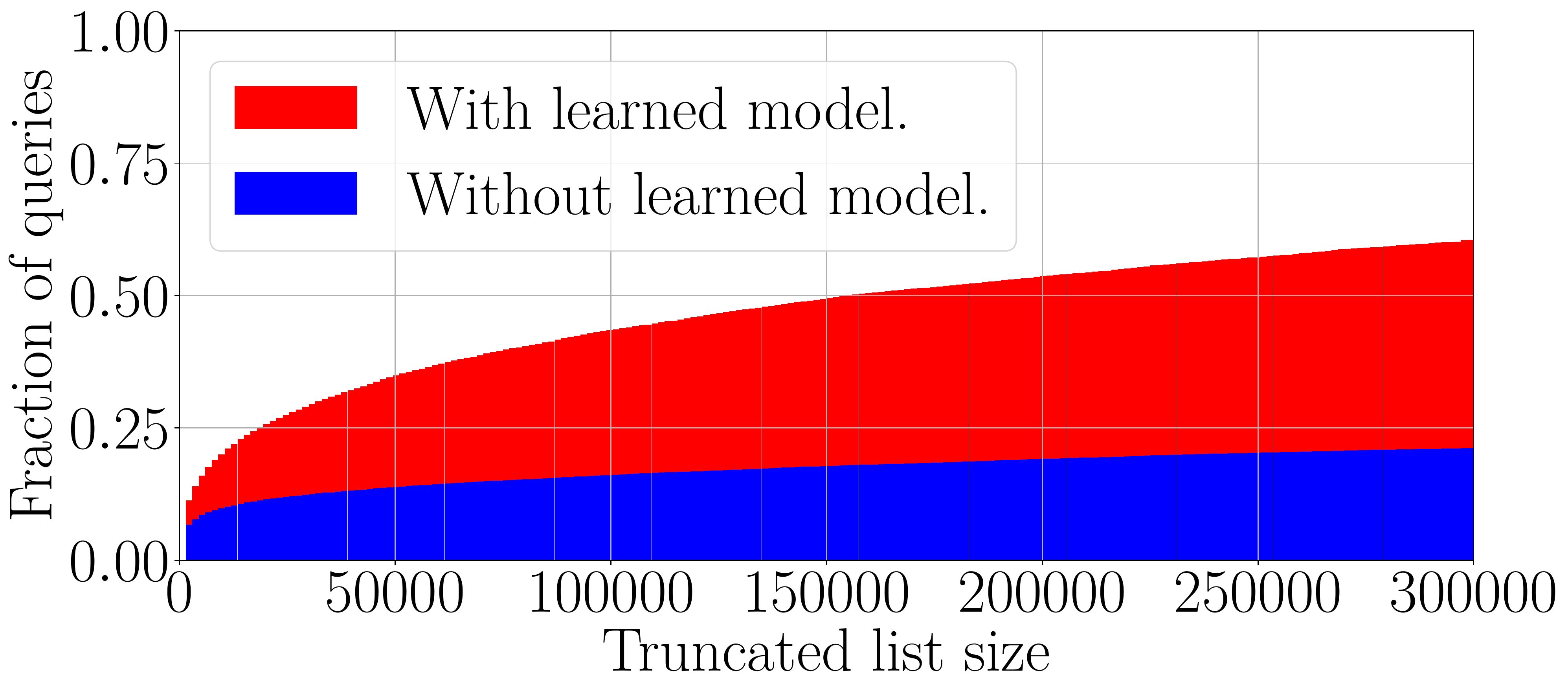}
\label{fig:affected}
\vspace{-1\baselineskip}
\end{figure*}

Finally, we conclude our answer to \ref{rq:possible}: there are
several methods by which a learned index structure could be applied
to Boolean intersection.
These approaches all make different tradeoffs between computational
costs during retrieval and gains in the amount of storage space required.
It may depend on the requirements of an application which approach is
the most suitable.

% !TEX root = adcs2018-potential-learned-indexing.tex

\section{Estimating Potential Gains}
\label{sec:gain}

In the previous section we proposed several approaches to support
conjunctive Boolean search with learned index structures.
In this section we will answer \ref{rq:gain} by estimating the gains
these approaches could make in terms of storage requirements.

For this analysis we will consider the two-tiered approach detailed
in Section~\ref{sec:topdoc}.
This approach was chosen because it appears to produce the least
storage gains, thus serving as a conservative bound, and, furthermore, for
this approach we can accurately estimate the tradeoffs it makes.
Three commonly used TREC document collections are considered for this
study: Robust 2005 (Newswire), GOV2, and
ClueWeb09B.\footnote{\url{https://trec.nist.gov}}
Figure~\ref{fig:freqcompr} displays the distribution of
document-frequencies in each collection.
Additionally, to see the varying storage terms require, it also shows
the minimum number of terms that can be stored in different fractions
of the compressed inverted index.
For this study we used OptPFOR compression~\cite{lemire2015decoding}.
From this figure it is clear that very few terms have a high document-frequency but that they can require a considerable percentage of
total storage cost in the inverted index.
For instance, in every collection we see that less than one percent
of the terms take up forty percent of storage.

To estimate the gains of the two-tiered approach we will use
truncated lists of a fixed size $k$ in the first partition.
Thus only terms with a higher document-frequency than $k$ will have truncated lists.
In addition this affects the optimization of $f$ as it only has to
consider terms for which not all documents are stored.
The potential gains in storage of the first partition are then estimated as follows: First,
we compute the amount of storage gained by removing the inverted
lists of replaced terms from the inverted index.
Second, we estimate the storage space required by a truncated list
of length $k$; we take the average size of compressed lists of the
same length in the complete compressed inverted
index~\cite{lemire2015decoding}.
Then we estimate the size of the learned model $f$ as linearly
proportional to the vocabulary and collection size: $|T|\cdot|D|\cdot
s$, where $s$ is an unknown positive value.
Lastly, we expect that for every term a bit has to be stored to
indicate whether it has been replaced or not.
By summing all these values we get the following formula for the
expected gain in storage; with $R$ as the set of terms to replace,
the complete set of terms $T$, and complete set of documents $D$:
\begin{align}
\begin{split}
\textit{ga}&\textit{in}(R, s) ={}\\
&\left[\sum_{t \in R} \textit{size.full.list}(t) - \textit{size.trunc.list}(k)\right] - (|R| + |D|) \cdot s - |T| .
\end{split}
\end{align}
To account for the unknown value of $s$ we compute a lower and upper
bound by varying its value.
For the upper bound, we estimate no cost from the model: $s = 0$,
this is the most gain this approach could potentially have.
The lower bound is estimated with $s = 512$ \emph{bits}, this is
equivalent to the cost of storing a compressed $128$ unit embedding for
every document and for every term as well.
We expect this to be the worst-case scenario in terms of model size.

Figure~\ref{fig:bounds} displays the estimated bounds for varying
truncated list sizes; in addition, it also shows the number of terms
that have to be replaced.
For instance, on the Robust collection, a gain of at least $40$\% can
be achieved by using a truncated list of \numprint{4000} and
replacing less than \numprint{4000} term lists.
Interestingly, the number of terms to replace grows exponentially as
the truncated list size decreases, while the potential gain
increases at a much smaller rate.
This further shows that the highest gains can be made by replacing
the most frequent terms.
Moreover, replacing extremely rare terms could even require more
storage depending on the model costs.
Regardless, we see that even with high model costs substantial gains
are possible by choosing an appropriate truncated list size.

Lastly, on a set of \numprint{40000} queries from the TREC Million
Query Track~\cite{carterette2009million},
we verified the number of queries with results that can be
guaranteed correct on the first partition.
\emph{With} a learned model a query is guaranteed correct if the list of \emph{at least one} term is not truncated.
In contrast, \emph{without} a learned model \emph{all} query-terms need complete lists for guaranteed correct results.
Figure~\ref{fig:affected} displays the difference between the two-tiered approach \emph{with} and \emph{without} the learned model.
As expected, the learned model considerably increases the correctness
of the results in the first stage.

Finally, we answer \ref{rq:gain} positively: our results show that
even the most storage inefficient approach with high model costs can
produce substantial reductions in storage requirements.

%\begin{figure*}[tb]
%\centering
%\includegraphics[width=0.32\textwidth]{img/robust/bounds100000}
%\includegraphics[width=0.32\textwidth]{img/gov2/bounds100000}
%\includegraphics[width=0.32\textwidth]{img/clueweb/bounds100000}\\
%\includegraphics[width=0.32\textwidth]{img/robust/terms_to_keep100000}
%\includegraphics[width=0.32\textwidth]{img/gov2/terms_to_keep100000}
%\includegraphics[width=0.32\textwidth]{img/clueweb/terms_to_keep100000}
%\caption{estimated bounds}
%\label{fig:long}
%\end{figure*}

% !TEX root = adcs2018-potential-learned-indexing.tex

\section{Conclusion}
\label{sec:conclusion}

In this study, we have explored how search based on Boolean
intersection may benefit from the usage of learned index structures.
We have proposed several approaches by which a learned model can
produce substantial reductions in storage requirements.
Each approach makes a tradeoff between storage requirements and
computational costs.
Our results show that even conservative estimates on the potential
gains w.r.t.\ space benefits are considerable.
We expect that combining learned index structures with inverted
indexes will be a fruitful research direction in the near future.

% I don't think it's worth effort to open source the code for this short paper.
%\subsection*{Code}
%To facilitate reproducibility of the results in this paper, we are sharing the code used to run the experiments in this paper at \\ \url{https://github.com/HarrieO/OnlineLearningToRank}.
%\vspace{-0.3\baselineskip}

%\subsection*{Acknowledgements}
%This research was partially supported by the Netherlands Organisation for Scientific Research (NWO) under project nr.\ 612.\-001.\-551.

%\begin{spacing}{1}
%\medskip\noindent\small
%\textbf{Acknowledgments.}
%%
%This research was supported by
%%
%Ahold Delhaize,
%%
%Amsterdam Data Science,
%%
%the Bloomberg Research Grant program,
%%
%the Criteo Faculty Research Award program,
%%
%the Dutch national program COMMIT,
%%
%Elsevier,
%%
%the European Community's Seventh Framework Programme (FP7/2007-2013) under
%grant agreement nr 312827 (VOX-Pol),
%%
%the Microsoft Research Ph.D.\ program,
%%
%the Netherlands Institute for Sound and Vision,
%%
%the Netherlands Organisation for Scientific Research (NWO)
%under pro\-ject nrs
%%
%612.\-001.\-116, % ImFIRE
%HOR-11-10, % HORIZON
%CI-14-25, % MediaNow
%652.\-002.\-001, % Re-Search
%612.\-001.\-551, % CLEAR
%652.\-001.\-003, % NEWEL
%%
%and
%%
%Yandex.
%%
%All content represents the opinion of the authors, which is not necessarily shared or endorsed by their respective employers and/or sponsors.
%\end{spacing}
%
%\vspace*{-.5\baselineskip}

%\balance

\myparagraph{Acknowledgements}
This work was partially supported by the Netherlands Organisation for Scientific Research (NWO) under project nr.\ 612.\-001.\-551 and
the Australian Research Council's {\emph{Discovery Projects}} Scheme (DP170102231). 

\balance

\bibliographystyle{ACM-Reference-Format}
\bibliography{strings-shrt,adcs2018-potential-learned-indexing}

\end{document}